\documentclass[conference, a4paper]{IEEEtran}
\IEEEoverridecommandlockouts
\usepackage{cite}
\usepackage{amsmath,amssymb,amsfonts}
\usepackage{algorithmic}
\usepackage{graphicx}
\usepackage{textcomp}
\usepackage{xcolor}
\usepackage{fancyhdr}
\usepackage{booktabs}    
\usepackage{comment}
\usepackage{multirow}    
\usepackage{graphicx}
\usepackage{subcaption} 
\usepackage{colortbl, xcolor}
\usepackage{float}
\usepackage{placeins}
\usepackage[T1]{fontenc}
\usepackage{lmodern}
\usepackage{setspace}
\def\BibTeX{{\rm B\kern-.05em{\sc i\kern-.025em b}\kern-.08em
    T\kern-.1667em\lower.7ex\hbox{E}\kern-.125emX}}
\begin{document}

\title{Cross-Scale Performance Analysis of Metaheuristic Algorithms for Simultaneous DG and DSTATCOM Placement in Radial Distribution Networks
}

\author{\IEEEauthorblockN{1\textsuperscript{st} Md. Tanvirul Islam}
\IEEEauthorblockA{\textit{Department of Electrical and Electronic Engineering} \\
\textit{Bangladesh University of Engineering and Technology}\\
Dhaka, Bangladesh \\
Email: 2006095@eee.buet.ac.bd}
}

\maketitle

\begin{abstract}
The problem of simultaneous placement of distributed generators and DSTATCOMs in radial distribution networks (RDNs) is a combinatorial mixed-integer optimization problem whose scalability with growing decision dimensionality has been insufficiently explored. A cross-scale analysis of seven metaheuristic algorithms, GWO, SCA, PSO, WOA, GA, HHO and SMA, is conducted on the IEEE 33-bus, 69-bus, and 136-bus systems at three problem dimensions \( d = 4, 8, 12 \), with 30 independent runs per configuration being validated through Wilcoxon and Friedman tests. Mean-performance statistics are extended with a Catastrophic Failure Rate (CFR) metric. The main result will be that dimensional scaling serves as a behavioral discriminator. The Friedman \( \chi^2 \) rises with the dimensionality, reaching its maximum value of \( \chi^2 = 143.79 \) in the 136-bus at \( d = 12 \) that corresponds to the progressive phase separation of the algorithms into two clusters of high and low performance. GA is the best performer in terms of the lowest rank in all the configurations. SCA has low variance but convergence to increasingly sub-optimal solutions. HHO exhibits catastrophic instability at all scales. Perhaps most strikingly, GA and PSO obtain a \( 3.3\% \) CFR on the 136-bus at \( d = 12 \) while the 33-bus at identical dimensionality has a \( 73\%-83\% \) CFR, showing that topology influences the reliability in a manner that is not captured by single-system measures.

\end{abstract}

\begin{IEEEkeywords}
Distributed Generation, DSTATCOM, Radial Distribution Network, Metaheuristic Optimization, Cross-Scale Comparative Analysis, Catastrophic Failure Rate
\end{IEEEkeywords}

\section{Introduction}
Active power losses continue to be a major operational issue in radial distribution systems. For example, in the IEEE 33-bus benchmark network, base-case active power losses are approximately 210\,kW, corresponding to about 5.7\% of the total load~\cite{b1}. Distributed generators (DGs) reduce upstream $I^2R$ losses by 40-70\% in benchmark studies~\cite{b2}, while DSTATCOMs provide reactive power support that improves voltage profiles within the $\pm 5\%$ limits of IEEE Std.~1547~\cite{b3}.

However, the effectiveness of both mentioned technologies strongly depends on their placement and sizing. Suboptimal siting may reduce expected benefits, but it may introduce new constraint violations~\cite{b4}. The simultaneous optimization of multiple DGs and DSTATCOMs is a mixed-integer nonlinear programming problem, where location variables are discrete and device sizes are continuous within operational limits. For the IEEE 136-bus system, the location space alone yields approximately $6.33 \times 10^{12}$ possible configurations for three DGs and three DSTATCOMs, making comprehensive search computationally intractable.

Several metaheuristic optimization techniques have been applied to this problem individually. A key question is whether algorithm ranking remains stable as dimensionality increases or changes when multiple devices are optimized simultaneously. This is particularly important in distribution planning, since an algorithm performing well for single-device cases may not scale effectively to multi-device scenarios. Therefore, the proper understanding of these scalability characteristics is essential for reliable practical scenarios.

This paper addresses this gap by systematically evaluating seven metaheuristic algorithms across multiple problem dimensions and IEEE radial distribution systems. The main contributions are summarized as follows:

\begin{itemize}
\item A scalable optimization framework for simultaneous placement and sizing of $N_{DG}$ distributed generators and $N_{DST}$ DSTATCOMs, enabling consistent evaluation across problem dimensions $d = 4, 8, 12$.
\item A comparative study of GWO, SCA, PSO, WOA, GA, HHO, and SMA on IEEE 33-bus, 69-bus, and 136-bus systems, demonstrating the variance of algorithm rankings with system size.
\item Introduction of the Catastrophic Failure Rate (CFR) as a fitness metric that captures convergence failures often hidden by mean-based performance measures in distribution system optimization.
\end{itemize}

\section{Literature Review}

\subsection{DG Siting in Distribution Networks}

The earliest analytical methods computed sensitivity indices of losses based on the DistFlow power flow equations~\cite{b1}. Moradi and Abedini employed a hybrid GA-PSO algorithm for multi-DG placement to reduce losses in IEEE 33-bus and 69-bus systems~\cite{b5}. Using the IEEE 33-bus and 136-bus systems, Devabalaji and Ravi applied a BF optimizer for simultaneous placement of DGs and DSTATCOMs~\cite{b8}. A common limitation in this literature is the use of a single system size for algorithm evaluation, with no assessment of scalability as the number of devices increases.

\subsection{DSTATCOM Placement on Distribution Networks}

Several analytical sensitivity-based methods have been proposed for single-DSTATCOM sizing under the assumption of fixed voltage profiles; however, they neglect nonlinear voltage-current coupling, which is critical for realistic loaded conditions~\cite{b4}. Selim et al. proposed a multi-objective SCA approach for optimal DSTATCOM placement in IEEE 33-bus and 69-bus systems, considering power loss, voltage deviation, and voltage stability index simultaneously~\cite{b9,b11}.

\subsection{Simultaneous DG and DSTATCOM Placement}

Devi and Geethanjali applied PSO for simultaneous placement in several test systems, showing that multiple active and reactive compensation devices are more effective than single-device configurations~\cite{b3}. Yuvaraj et al. used IEEE 33-bus and 136-bus systems for simultaneous DG and DSTATCOM allocation, providing reference results for large-network cases considered in this study~\cite{b10}. However, the impact of multiple devices per type (e.g., two or three) remains largely unexplored, as prior simultaneous placement studies typically consider only a single level of complexity.

\section{System Model and Problem Formulation}

\subsection{Distribution Network Model}

A radial distribution system can be represented as a directed graph $G(N, B)$, where $N$ is the set of buses and $B$ is the set of branches. Bus 1 is the substation (slack bus) with fixed voltage at 1.0\,pu. The remaining buses represent load points and potential locations for DG or DSTATCOM installation.


Power flow in each branch is governed by the DistFlow equations of Baran and Wu \cite{b1}. Network solution uses the Backward-Forward Sweep (BFS) method. Convergence is declared when the maximum per-unit voltage change between successive iterations falls below $10^{-6}$ pu.
Total active power loss is computed from converged branch flows,
\begin{equation}
    P_{loss} = \sum_{(i,j) \in B} \frac{R_{ij} * (P_{ij}^2 + Q_{ij}^2)}{V_i^2}   
\end{equation}
\vspace{-4pt}

\subsection{Test Systems}

Three radial IEEE distribution test systems are used, solved using a single-phase positive-sequence BFS load flow. The study considers the IEEE 33-bus and 69-bus systems at 12.66 kV, and the 136-bus system at 11 kV using the full network data from~\cite{b8}.

\vspace{-4pt}
\subsection{Decision Variables and Problem Configurations}

The solution vector for $N_{DG}$ distributed generators and $N_{DST}$ DSTATCOMs is, 
\begin{equation}
\begin{aligned}
    x = [b_{DG_1}, S_{DG_1}, b_{DG_2}, S_{DG2}, ..., b_{DG_m}, S_{DG_m}, \\
     b_{DST_1}, Q_{DST_1}, b_{DST_2}, Q_{DST_2}, ..., b_{DST_p}, Q_{DST_p}]
\end{aligned}
\end{equation}

where $b_{DG_k}$ is the 1-indexed bus location of DG k (integer), $S_{DG_k}$ is its active power rating (kW, continuous), $b_{DST_k}$ is the bus location of DSTATCOM k (integer), and $Q_{DST_k}$ is its reactive power rating (kVAR, continuous). The solution vector has dimension $d = 2*(N_{DG} + N_{DST})$.\\
This study evaluates three configurations of increasing complexity,\\
\textbf{Case 1:} $N_{DG}$ = 1, $N_{DST}$ = 1,  d = 4\\
\textbf{Case 2:} $N_{DG}$ = 2, $N_{DST}$ = 2,  d = 8\\
\textbf{Case 3:} $N_{DG}$ = 3, $N_{DST}$ = 3,  d = 12\\
\vspace{-4pt}
\subsection{Objective Function}

A composite weighted objective is minimized,
 \begin{equation}
     F(x) = W_1 * F_1(x) + W_2 * F_2(x) + W_3 * F_3(x)        
 \end{equation}
where $W_1$ = 0.5, $W_2$ = 0.3, $W_3$ = 0.2 are weights reflecting the primary engineering priority of loss reduction, with secondary and tertiary priorities assigned proportionally; sensitivity to these values is limited since all three sub-objectives are normalized to comparable magnitudes 

$F_1$ is the Power Loss Index, normalized to the base-case loss,
\begin{equation}
    F1(x) = \frac{P_{loss}(x)}{P_{loss}^{base}}                   
\end{equation}
$F_2$ is the Voltage Deviation Index, measuring the mean squared deviation of all non-slack bus voltages from nominal,
\begin{equation}
    F_2(x) = \frac{1}{(n-1)}  * \sum_{k=2}^{n} (1 - V_k)^2     
\end{equation}

$F_3$ is the normalized annual cost, comprising amortized capital expenditure and ongoing energy cost of residual losses,
\begin{equation}
\begin{aligned}
F_3(x) = & \frac{C_{DG} \sum(S_{DG_k})}{T_{life}} \\
        & + \frac{C_{DST} \sum(Q_{DST_k})}{T_{life}} \\
        & + \frac{C_{loss} \, P_{loss}(x)\, T_{yr}}{C_{base}}
\end{aligned}
\end{equation}
where $C_{DG} = 400~\$/\mathrm{kW}$, $C_{DST} = 50~\$/\mathrm{kVAR}$, $C_{loss} = 0.06~\$/\mathrm{kWh}$, $T_{yr} = 8760~\mathrm{h/yr}$, $T_{life} = 20~\mathrm{yr}$ (asset lifetime for capital amortization), and $C_{base} = C_{loss} \cdot P_{loss}^{base} \cdot T_{yr}$ is the base-case annual loss-energy cost used as the normalizer.

\subsection{Constraints and Penalty Handling}

Bus voltages are constrained within the range of \(0.95\text{-}1.05\,\mathrm{pu}\) in accordance with IEEE Std.~1547. DG active power output is limited to \(100\text{-}2500\,\mathrm{kW}\), while DSTATCOM reactive power injection is bounded between \(100\text{-}2000\,\mathrm{kVAR}\). DG units are assumed to operate at unity power factor, whereas DSTATCOMs provide purely reactive compensation. Any constraint violation is incorporated into the objective function through a quadratic penalty formulation.
\begin{equation}
\begin{aligned}
F_{pen}(x) = \; F(x) + \rho \sum_{k=1}^{n} \Big[ 
& \max(0, 0.95 - V_k)^2 \\
+ & \max(0, V_k - 1.05)^2
\Big]
\end{aligned}
\end{equation}

with $\rho = 10^4$. Non-convergent load-flow evaluations return a sentinel fitness of 10¹² and are immediately discarded. 

\subsection{Compared Algorithms}

To ensure a fair comparison, seven population-based metaheuristic algorithms are tested under identical settings without algorithm-specific parameter tuning. The population size is fixed at $N=30$ and the number of iterations at $T=200$, eliminating computational bias and emphasizing behavioral scalability.

GWO distinguishes agents into $\alpha$, $\beta$, $\delta$, and $\omega$ ranks, where the three best solutions guide the search and the control parameter $a$ decreases linearly from 2 to 0 \cite{b7}.

SCA updates positions using sinusoidal functions towards the current best solution~\cite{b14}, with the attenuation factor
\begin{equation}
r_1(t) = 2 \left(1 - \frac{t}{T}\right),
\end{equation}
which decays to zero over iterations, progressively reducing exploration.

PSO updates particle velocities using cognitive and social components~\cite{b16}, with linearly decreasing inertia $(0.9 \to 0.4)$, $c_1=c_2=2.0$, and velocity clamping at $\pm0.2(ub-lb)$.

WOA alternates between encircling prey, spiral attack, and random search, governed by
$A_s = 2a \cdot \mathrm{rand} - a$~\cite{b15}.

GA uses tournament selection $(k=3)$, SBX crossover $(\eta_c=20)$, polynomial mutation $(\eta_m=20, p_{\mathrm{mut}}=0.1)$, and elitism~\cite{b17}, with post-operator rounding to enforce integer decision variables.

HHO models exploration and exploitation via escape energy, where the transition depends on time-varying energy dynamics, enabling switches between exploration and four attack modes~\cite{b18}.

SMA sorts agents by fitness at each iteration and applies adaptive weights that decay toward zero as $t \to T$, ensuring convergence~\cite{b19}.

 \subsection{Statistical Evaluation Framework}

\subsubsection{Performance Metrics}

There are 30 independent runs per configuration per test system for each algorithm. Two performance metrics are reported.

\textbf{Mean fitness ($\mu$):} The average best fitness over 30 trials, representing typical solution quality.

\textbf{Catastrophic Failure Rate (CFR):} The percentage of runs in which the final best fitness exceeds twice the best-known fitness for a given case. Formally, for a set of 30 trial results $\{f_1, \dots, f_{30}\}$,
\begin{equation}
\mathrm{CFR} = \frac{1}{30} \sum_{i=1}^{30} \mathbf{1}\left[f_i > 2 f_{\min}\right] \times 100\%
\label{eq:cfr}
\end{equation}
where $f_{\min}$ is the minimum fitness value obtained across all algorithms and trials for that configuration and test system, and $\mathbf{1}[\cdot]$ is an indicator function. The factor $2\times$ is chosen as the smallest threshold that consistently separates operationally undesirable solutions from typical stochastic variation in clustered placement objectives.

\subsubsection{Statistical Hypothesis Tests}

The Wilcoxon signed-rank test is used for pairwise comparison of algorithms with a one-tailed alternative hypothesis, and results are considered statistically significant for $p < 0.05$. For overall comparison, the Friedman rank test is applied, where mean ranks are computed across trials and tested for equality using a chi-squared statistic at the 0.05 significance level.

\section{Results and Discussion}
A set of 30 independent trials is conducted for each base case ($N=30$, $T=200$), and results are compared against reference values (Table~\ref{tab:test_systems}). All solutions satisfy the IEEE STD. 1547 voltage limits ($\pm5\%$). Dimensional scaling highlights differences in exploration and convergence behavior across algorithms. The $2\times$ CFR metric captures practical non-competitiveness by identifying solutions that deviate from near-optimal performance. Loss reduction ranges from 65-97\% for the 33- and 69-bus systems and 29-51\% for the 136-bus system, depending on load distribution.

    \label{lossbar}

\begin{table}[!t]
\caption{Test System Characteristics}
\label{tab:test_systems}
\centering
\footnotesize
\setlength{\tabcolsep}{3pt}
\renewcommand{\arraystretch}{1.05}

\begin{tabular}{cccccc}
\toprule
\textbf{System} & \textbf{Bus} & \textbf{Branch} &
\textbf{$V_{\mathrm{base}}$} &
\textbf{$P_{\mathrm{loss}}$} &
\textbf{$V_{\min}$} \\

& & & \textbf{(kV)} & \textbf{(kW)} & \textbf{(pu)} \\
\midrule

33-bus  & 33  & 32  & 12.66 & 202.68 & 0.9131 \\
69-bus  & 69  & 68  & 12.66 & 223.21 & 0.9092 \\
136-bus & 136 & 135 & 11.00 & 320.39& 0.9307    \\

\bottomrule
\end{tabular}
\end{table}

\subsection{Baseline Validation (Case 1, $d = 4$)}

CFR values show meaningful reliability structure even at the smallest problem scale. At $d=4$, HHO already exhibits high failure rates exceeding 60\% across all systems. On the standard benchmarks, GWO, WOA, and PSO show moderate CFR, while GA and SCA are the most reliable. The Friedman test indicates statistically significant differences among all systems ($p < 0.001$, Table~\ref{tab:stats}), with GA achieving the best mean rank and HHO consistently the worst. Notably, SCA attains the best single-run loss reduction on the 69-bus system but ranks fifth in Friedman analysis; its near-zero inter-run variance indicates convergence to a sub-optimal basin rather than robust optimization. This behavior is linked to its deterministic step-size schedule.

\begin{table}[!t]
\caption{ \\ Mean fitness ($\mu$) and Friedman mean rank (Fr.R) for each algorithm.\\
$^{***}p<0.001$, $^{**}p<0.01$, $^{*}p<0.05$,
$^{ns}$: not significant. \\Bold indicates best $\mu$.}
\label{tab:stats}

\centering
\scriptsize
\setlength{\tabcolsep}{2.6pt}
\renewcommand{\arraystretch}{1.03}

\begin{tabular}{clcccccc}
\toprule

&
&
\multicolumn{2}{c}{\textbf{33-bus}} &
\multicolumn{2}{c}{\textbf{69-bus}} &
\multicolumn{2}{c}{\textbf{136-bus}} \\

\cmidrule(lr){3-4}
\cmidrule(lr){5-6}
\cmidrule(lr){7-8}

\textbf{Alg} & \textbf{C}
& $\mu$ & Fr.R
& $\mu$ & Fr.R
& $\mu$ & Fr.R \\

\midrule

\multirow{3}{*}{GWO}
& C1 & $0.2745^{ns}$ & 3.33 & $0.1447^{***}$ & 3.90 & {$0.5348^{ns}$} & {4.1} \\
& C2 & $0.1830^{ns}$ & 3.17 & $0.1186^{***}$ & 3.87 & {$0.5099^{ns}$} & {4.13} \\
& C3 & $0.1600^{*}$  & 3.00 & $0.1131^{***}$ & 3.87 & $0.4858^{***}$ & 4.17 \\

\midrule

\multirow{3}{*}{SCA}
& C1 & $0.2706^{ns}$ & 3.90 & $0.1343^{***}$ & 5.27 & {$0.5328^{ns}$} & {4.20} \\
& C2 & $0.1824^{*}$  & 3.70 & $0.1221^{***}$ & 4.87 & {$0.5116^{ns}$} & {4.63} \\
& C3 & $0.1670^{**}$ & 4.30 & $0.1213^{***}$ & 5.07 & $0.4965^{***}$ & 5.17 \\

\midrule

\multirow{3}{*}{PSO}
& C1 & $0.2809^{ns}$ & 3.03 & $0.1653^{ns}$ & 2.60 & {$0.5385^{ns}$} & {2.53} \\
& C2 & $0.1887^{ns}$ & 3.00 & $0.1155^{ns}$ & 1.97 & {$0.4948^{ns}$} & {1.20} \\
& C3 & $0.1591^{*}$  & 2.63 & $0.0988^{ns}$ & 2.20 & $0.4663^{*}$ & 1.93 \\

\midrule

\multirow{3}{*}{WOA}
& C1 & $0.2774^{ns}$ & 2.67 & $0.1570^{ns}$ & 2.17 & {$0.5394^{ns}$} & {3.57} \\
& C2 & $0.1897^{**}$ & 3.63 & $0.1110^{ns}$ & 2.87 & {$0.5035^{ns}$} & {3.43} \\
& C3 & $0.1651^{**}$ & 3.50 & $0.1044^{***}$ & 2.97 & $0.4752^{***}$ & 3.13 \\

\midrule

\multirow{3}{*}{\textbf{GA}}
& C1 & $\mathbf{0.2727}$ & $\mathbf{2.97}$
     & $\mathbf{0.1324}$ & $\mathbf{2.20}$
     & {$\mathbf{0.5314^{ns}}$} & {$\mathbf{2.2}$} \\
& C2 & $\mathbf{0.1821}$ & $\mathbf{2.43}$
     & $\mathbf{0.1056}$ & $\mathbf{1.73}$
     & $\mathbf{0.4913^{ns}}$ & $\mathbf{1.73}$ \\
& C3 & $\mathbf{0.1566}$ & $\mathbf{2.10}$
     & $\mathbf{0.0969}$ & $\mathbf{1.70}$
     & $\mathbf{0.4619}$ & $\mathbf{1.40}$ \\

\midrule

\multirow{3}{*}{HHO}
& C1 & $0.3025^{***}$ & 6.87 & $0.1967^{***}$ & 6.93 & {$0.5558^{***}$} & {6.70} \\
& C2 & $0.2357^{***}$ & 6.83 & $0.1878^{***}$ & 6.83 & {$0.5332^{***}$} & {6.57} \\
& C3 & $0.2325^{***}$ & 6.83 & $0.2080^{***}$ & 7.00 & $0.5178^{***}$ & 6.53 \\

\midrule

\multirow{3}{*}{SMA}
& C1 & $0.2759^{***}$ & 5.23 & $0.1332^{***}$ & 4.67 & {$0.5365^{*}$} & {4.90} \\
& C2 & $0.1978^{***}$ & 5.23 & $0.1214^{***}$ & 5.60 & {$0.5192^{***}$} & {5.60} \\
& C3 & $0.1827^{***}$ & 5.63 & $0.1221^{***}$ & 5.20 & $0.5021^{***}$ & 5.67 \\

\specialrule{0.08em}{0.3ex}{0.3ex}

\multirow{3}{*}{\textit{Friedman}}
& C1 & \multicolumn{2}{c}{$89.79^{***}$}
     & \multicolumn{2}{c}{$117.89^{***}$}
     & \multicolumn{2}{c}{$88.47^{***}$} \\
& C2 & \multicolumn{2}{c}{$89.50^{***}$}
     & \multicolumn{2}{c}{$100.13^{***}$}
     & \multicolumn{2}{c}{$123.09^{***}$} \\
& C3 & \multicolumn{2}{c}{$112.59^{***}$}
     & \multicolumn{2}{c}{$136.24^{***}$}
     & \multicolumn{2}{c}{$143.79^{***}$} \\

\bottomrule
\end{tabular}
\end{table}



\definecolor{cfl}{gray}{0.88}
\definecolor{cfm}{gray}{0.72}
\definecolor{cfd}{gray}{0.45}
\definecolor{cfx}{gray}{0.20}

\begin{table}[!b]
\caption{Catastrophic Failure Rate (\%)}
\label{tab:cfr}
\centering
\scriptsize
\setlength{\tabcolsep}{3pt}
\renewcommand{\arraystretch}{1.12}
%
\newcommand{\wt}[1]{\textcolor{white}{\textbf{#1}}}
\begin{tabular}{@{}l ccc ccc ccc @{}}
\toprule
& \multicolumn{3}{c}{\textbf{IEEE 33-bus}}
& \multicolumn{3}{c}{\textbf{IEEE 69-bus}}
& \multicolumn{3}{c@{}}{\textbf{136-bus}} \\
\cmidrule(lr){2-4}\cmidrule(lr){5-7}\cmidrule(lr){8-10}
\textbf{Algo}
  & C1 & C2 & C3
  & C1 & C2 & C3
  & C1 & C2 & C3 \\
\midrule

GWO
  & \cellcolor{cfm}20.0
  & \cellcolor{cfm}16.7
  & \cellcolor{cfx}\textcolor{white}{96.7}
  & \cellcolor{cfm}16.7
  & \cellcolor{cfx}\textcolor{white}{93.3}
  & \cellcolor{cfx}\textcolor{white}{96.7}
  & \cellcolor{white}0.0
  & \cellcolor{cfm}33.3
  & \cellcolor{cfd}\textcolor{white}{56.7} \\

SCA
  & \cellcolor{white}0.0
  & \cellcolor{cfm}16.7
  & \cellcolor{cfx}\textcolor{white}{100.0}
  & \cellcolor{white}0.0
  & \cellcolor{cfx}\textcolor{white}{100.0}
  & \cellcolor{cfx}\textcolor{white}{100.0}
  & \cellcolor{white}0.0
  & \cellcolor{cfd}\textcolor{white}{60.0}
  & \cellcolor{cfx}\textcolor{white}{100.0} \\

PSO
  & \cellcolor{cfl}10.0
  & \cellcolor{cfm}30.0
  & \cellcolor{cfx}\textcolor{white}{83.3}
  & \cellcolor{cfm}33.3
  & \cellcolor{cfd}\textcolor{white}{56.7}
  & \cellcolor{cfl}{3.3}
  & \cellcolor{cfl}6.7
  & \cellcolor{white}0.0
  & \cellcolor{cfl}3.3 \\

WOA
  & \cellcolor{cfm}20.0
  & \cellcolor{cfm}26.7
  & \cellcolor{cfx}\textcolor{white}{83.3}
  & \cellcolor{cfm}16.7
  & \cellcolor{cfl}3.3
  & \cellcolor{cfd}\textcolor{white}{70.0}
  & \cellcolor{white}0.0
  & \cellcolor{cfl}16.7
& \cellcolor{cfd}\textcolor{white}{46.7} \\

GA
  & \cellcolor{cfl}6.7
  & \cellcolor{cfm}30.0
  & \cellcolor{cfx}\textcolor{white}{73.3}
  & \cellcolor{cfl}3.3
  & \cellcolor{cfd}\textcolor{white}{53.3}
  & \cellcolor{cfm}26.7
  & \cellcolor{white}0.0
  & \cellcolor{white}3.3
  & \cellcolor{white}3.3 \\

HHO
  & \cellcolor{cfd}60.0
  & \cellcolor{cfx}\textcolor{white}{100.0}
  & \cellcolor{cfx}\textcolor{white}{100.0}
  & \cellcolor{cfx}\textcolor{white}{83.3}
  & \cellcolor{cfx}\textcolor{white}{100.0}
  & \cellcolor{cfx}\textcolor{white}{100.0}
  & \cellcolor{cfd}\textcolor{white}{53.3}
  & \cellcolor{cfx}\textcolor{white}{90.0}
  & \cellcolor{cfx}\textcolor{white}{100.0} \\

SMA
  & \cellcolor{white}0.0
  & \cellcolor{cfx}\textcolor{white}{80.0}
  & \cellcolor{cfx}\textcolor{white}{100.0}
  & \cellcolor{white}0.0
  & \cellcolor{cfx}\textcolor{white}{96.7}
  & \cellcolor{cfx}\textcolor{white}{100.0}
  & \cellcolor{white}0.0
  & \cellcolor{cfx}\textcolor{white}{73.3}
  & \cellcolor{cfx}\textcolor{white}{100.0} \\

\bottomrule
\end{tabular}
\end{table}

%
%

\subsection{Dimensionality-Induced Rank Stratification (Case 2, \textit{d} = 8)}

Increasing dimensionality amplifies ranking separation among algorithms without necessarily causing uniform performance degradation. The Friedman $\chi^2$ statistics (Table~\ref{tab:stats}) increase monotonically on the 136-bus system (88.47 \(\to\) 123.09 \(\to\) 143.79 across Cases 1–3), the strongest scaling trend in the study, and also rise on the 69-bus system from 117.89 to 136.24. This reflects a phase-separation-like behavior, where algorithms increasingly cluster into high-performance (GA, PSO) and low-performance (SCA, SMA, HHO) groups as dimensionality increases.

\begin{figure}[!t]
    \centering
    \includegraphics[width=0.75\linewidth]{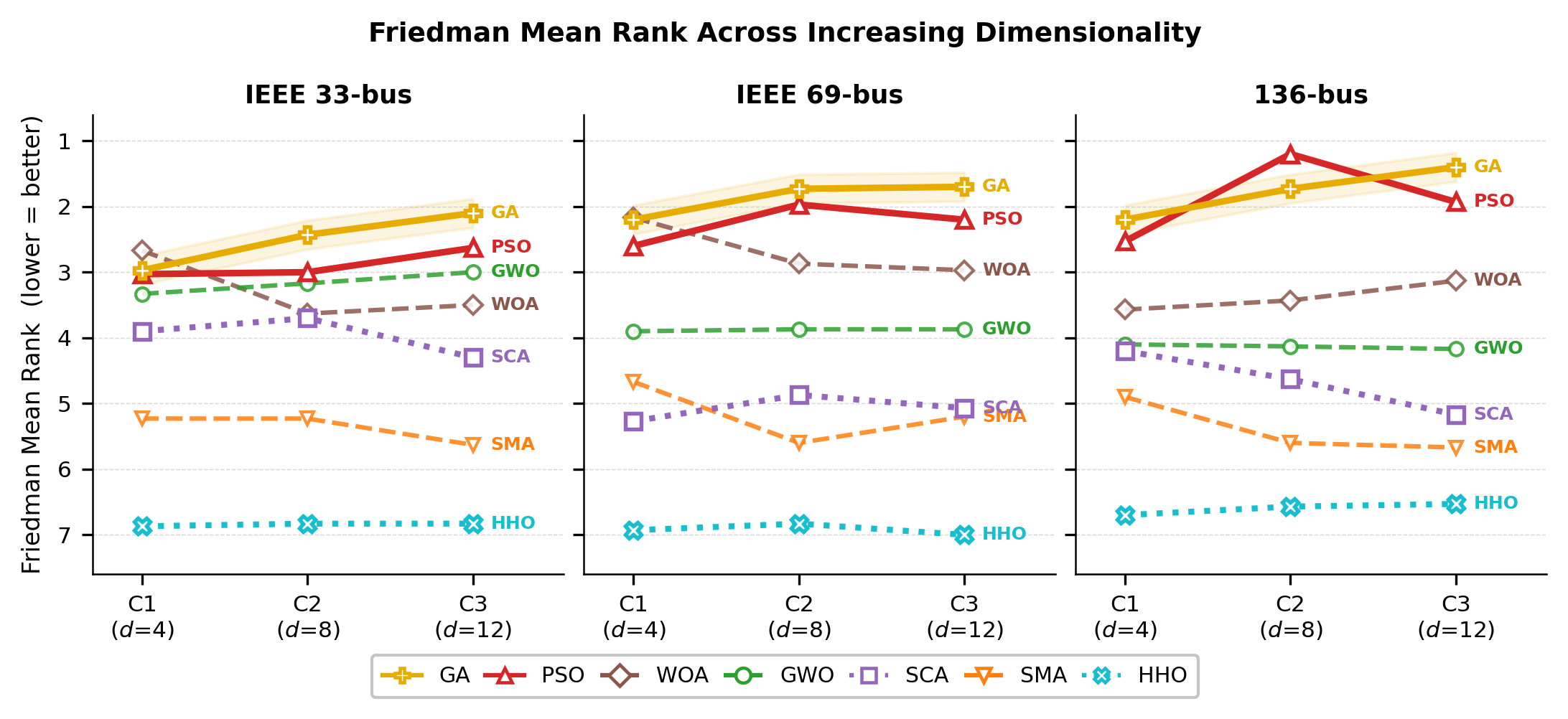}
    \caption{Friedman mean rank of each algorithm across increasing problem dimensionality on IEEE-33, 69- and 136-bus system}
    \label{fig:placeholder}
\end{figure}

At $d=8$, SCA, SMA, and HHO exhibit severe CFR escalation on the standard benchmarks, with HHO reaching 100\% in both cases (Table~\ref{tab:cfr}). GA and PSO also deteriorate but remain the most competitive. The trend differs on the 136-bus system, where PSO achieves a Friedman rank of 1.20, the lowest single-configuration rank in the study. In contrast, SCA and SMA already perform poorly on the same system. This topology-dependent divergence becomes more pronounced at $d=12$, as shown in Fig.~\ref{fig:placeholder}.

\subsection{Full-Scale Collapse and Mechanism (Case 3, \textit{d} = 12)}

Near-universal CFR is observed on the standard benchmarks at $d=12$, with SCA, HHO, and SMA reaching 100\% and GWO close behind. GA and PSO remain partially viable, although GA is non-competitive in nearly 75\% of runs on the 33-bus system. Simultaneously, Friedman $\chi^2$ values increase across all systems as reliability decreases, confirming greater statistical separation under lower performance regimes. Across all nine system--case configurations, GA achieves the best Friedman mean rank among all algorithms.

\begin{figure}[!t]
\centering

\begin{subfigure}{0.32\columnwidth}
    \centering
    \includegraphics[width=\linewidth]{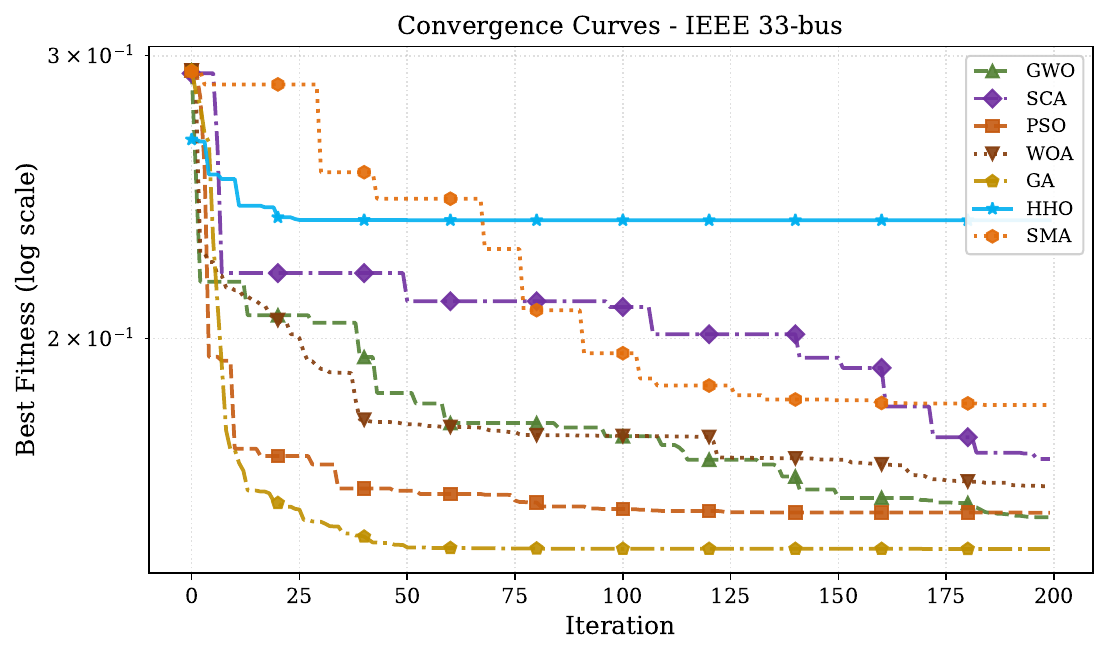}
    \caption{33-Bus System}
    \label{fig:sub_a1}
\end{subfigure}
\hfill
\begin{subfigure}{0.32\columnwidth}
    \centering
    \includegraphics[width=\linewidth]{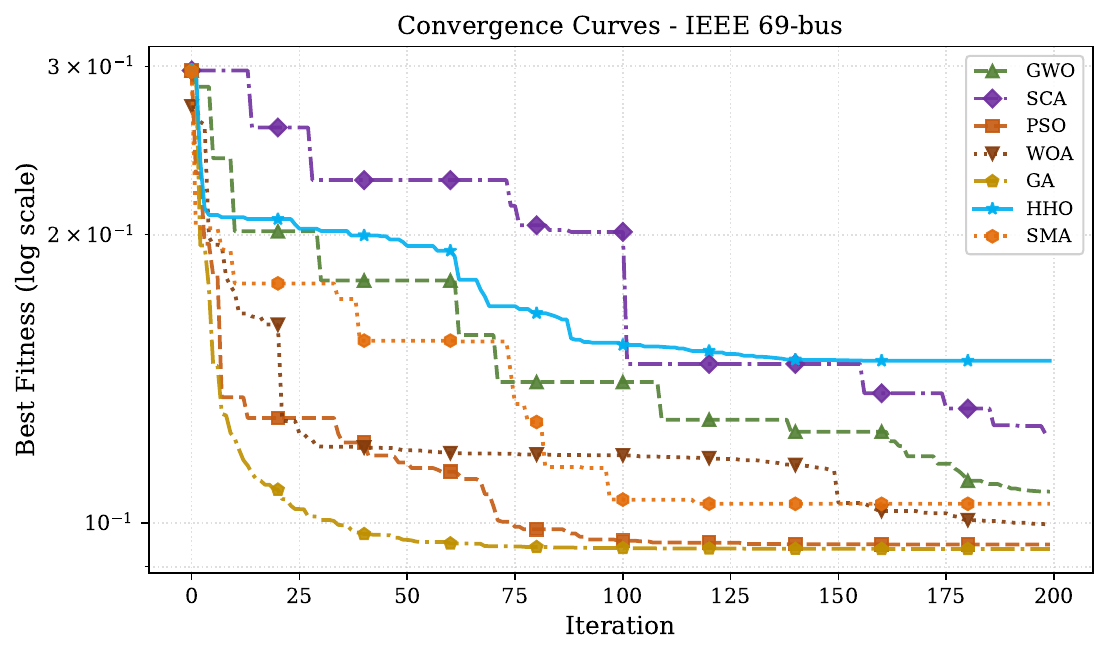}
    \caption{69-Bus System}
    \label{fig:sub_b1}
\end{subfigure}
\hfill
\begin{subfigure}{0.32\columnwidth}
    \centering
    \includegraphics[width=\linewidth]{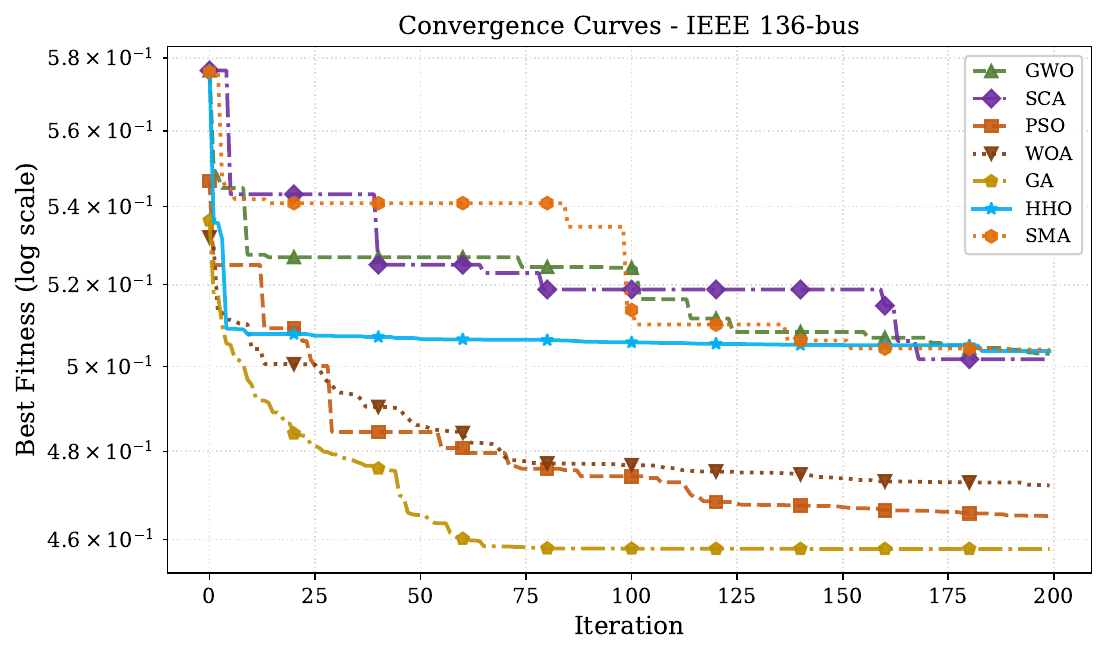}
    \caption{136-Bus System}
    \label{fig:sub_c1}
\end{subfigure}

\caption{Best-run convergence characteristics for Case~3.}
\label{fig:convergence}
\end{figure}

\begin{figure}[!t]
\centering

\begin{subfigure}{0.32\columnwidth}
    \centering
    \includegraphics[width=\linewidth]{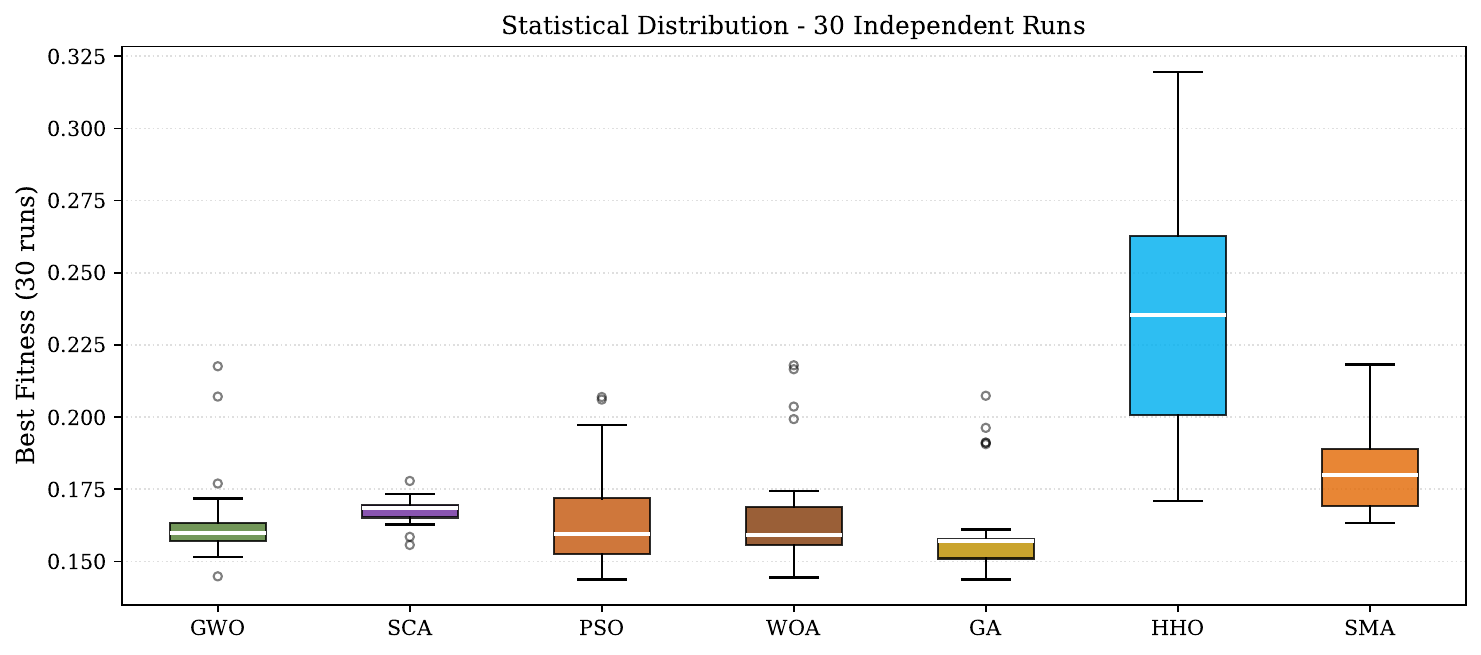}
    \caption{33-Bus System}
    \label{fig:sub_a2}
\end{subfigure}
\hfill
\begin{subfigure}{0.32\columnwidth}
    \centering
    \includegraphics[width=\linewidth]{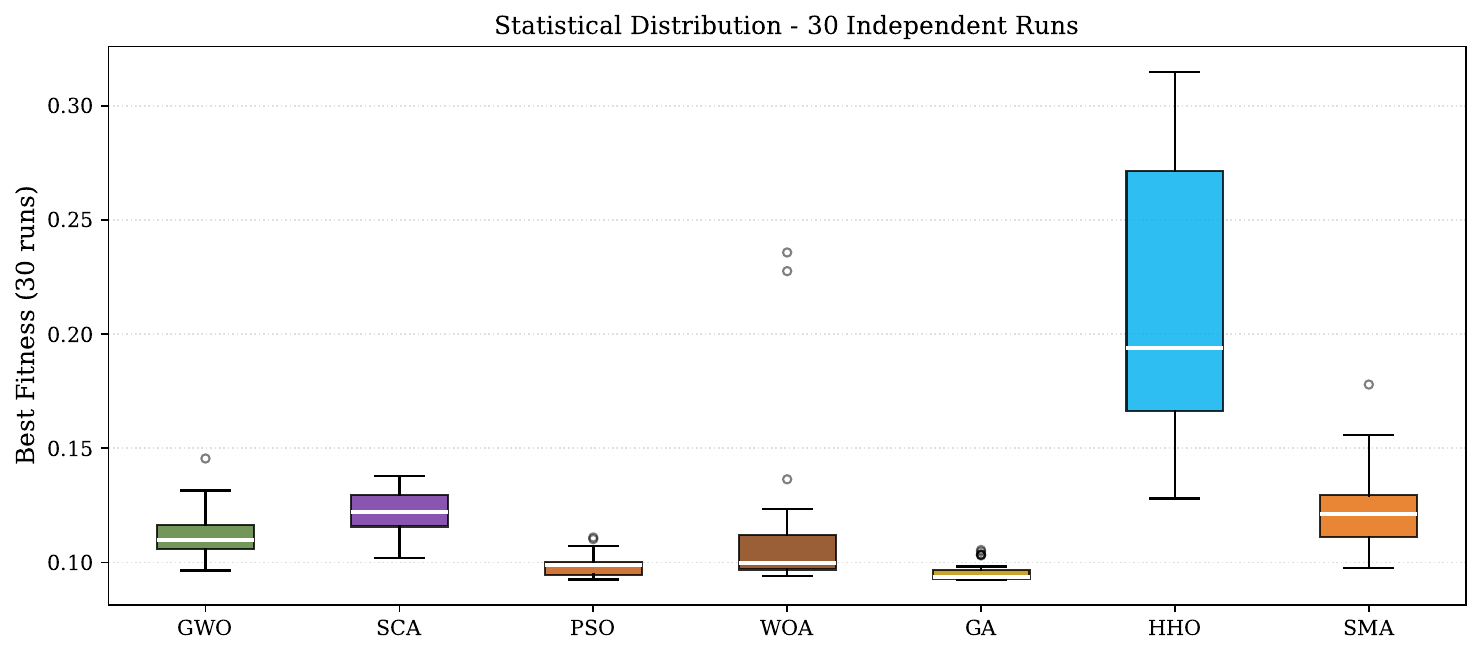}
    \caption{69-Bus System}
    \label{fig:sub_b2}
\end{subfigure}
\hfill
\begin{subfigure}{0.32\columnwidth}
    \centering
    \includegraphics[width=\linewidth]{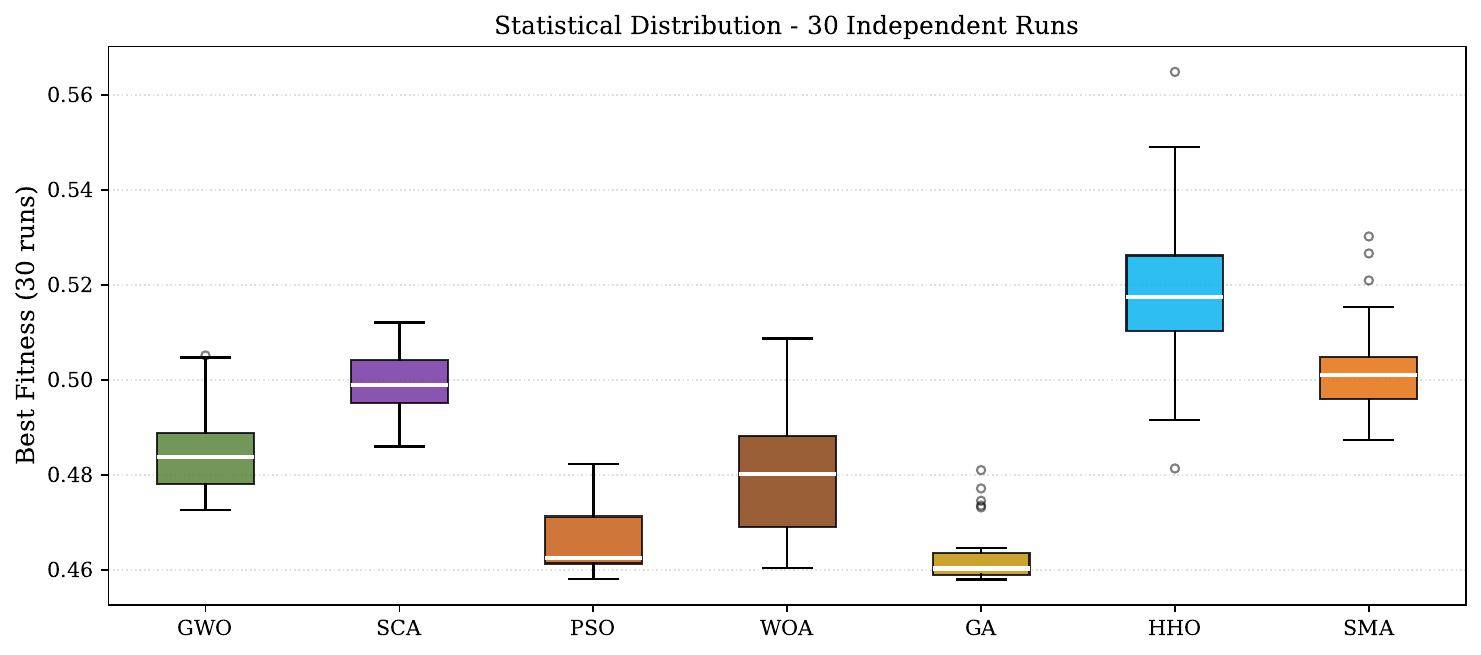}
    \caption{136-Bus System}
    \label{fig:sub_c2}
\end{subfigure}

\caption{30-run fitness distribution at Case~3.}
\label{boxplot}
\end{figure}

Fig.~\ref{fig:convergence} reveals the underlying optimization dynamics. GA and PSO exhibit rapid convergence toward low-fitness regions across all systems, whereas SCA and SMA stagnate within the first 50 iterations. HHO, in contrast, demonstrates oscillatory and unstable search trajectories. Fig.~\ref{boxplot} further confirms these behaviours from a distributional perspective. GA achieves the lowest median fitness with compact dispersion, while HHO exhibits elevated medians and wide interquartile ranges.

GA is able to maintain its scalability by introducing diversification at each step, via recombination, which produces structural variation to keep exploration going under the combinatorial expansion. PSO shows a similar adaptability in its cognitive-social memory structure. SCA uses a deterministic sinusoidal decay schedule, which reduces the size of the search space regardless of any fitness response, thereby resulting in premature convergence as the search space grows. The behaviour of SMA is similar to contraction-driven behaviour. The exploitation phase of HHO is not stable for all of the dimensional scales studied, leading to very irregular convergence properties.

 \subsection{Topology-Dependent Reliability Reversal}
The 136-bus system yields the most pronounced results. At $d=12$, GA and PSO achieve CFR values of 3.3\%, while SCA, HHO, and SMA reach 100\%, consistent with their performance on the 33-bus system under the same settings. It also exhibits the highest Friedman $\chi^2$ statistic (143.79) and the largest GA--SCA fitness separation. The improvement from Case~2 to Case~3 is greater on the 136-bus system (7.2\,pp vs. 0.5\,pp on the 69-bus), indicating a more exploitable fitness structure at higher dimensionality. This suggests that GA and PSO better exploit distributed regions of high-quality solutions due to crossover and social memory mechanisms. The effect is topology-driven: the 136-bus system has a higher bus-to-device ratio, producing a more dispersed search space and wider fitness basins. Consequently, conclusions drawn from smaller systems alone would misleadingly suggest uniform algorithm failure at $d=12$, whereas the 136-bus system reveals clear topology-dependent separability.

\subsection{Engineering Implications}

According to CFR, a catastrophic failure is defined as a trial in which the final fitness exceeds twice the world-wide best solution (i.e., a $2\times$ threshold). The ranking patterns remain qualitatively unchanged under alternative thresholds such as $1.5\times$ and $3\times$.

Three key insights emerge from this study. First, single-topology benchmarks are not scalable and may be misleading at low dimensions. Second, low inter-run variance does not necessarily indicate high-quality solutions; for example, the near-zero variance observed in SCA reflects consistent convergence to an inferior basin rather than robust optimization, highlighting the need to use CFR alongside rank-based metrics. Third, network topology significantly influences optimization reliability in ways not captured by traditional benchmark evaluations. Overall, scalability should be treated as a primary evaluation criterion in power-system metaheuristic research, in addition to performance at fixed system sizes.


\vspace{-4pt} 

\section{Conclusion}
This study shows that increasing dimensionality primarily amplifies ranking separation among metaheuristics rather than uniformly improving or degrading performance. The Friedman $\chi^2$ statistics reach 112.59, 136.24, and 143.79 at $d=12$, representing the highest values observed in this study. Although best-run performance improves, distributional quality deteriorates for algorithms such as SCA and SMA, which is captured more effectively through the CFR metric. Future work will incorporate explicit landscape measures and time-varying load profiles to further characterize scalability behavior.
\section*{Acknowledgment}

The author thanks the Department of Electrical and Electronic Engineering (EEE), Bangladesh University of Engineering and Technology(BUET), for its support. The author also acknowledges the use of \textit{Claude} (\textit{Anthropic}) for grammar refinement and text condensation, and \textit{OpenAI Codex} for code refinement assistance. The authors take full responsibility for all content presented in this
paper.

\end{document}